\documentclass[a4paper,10pt,twoside]{cpc-hepnp}
\usepackage{CJK,upgreek,fancyhdr}
\usepackage{subfigure}
\usepackage{multicol}
\usepackage{booktabs}
\usepackage{amsmath}

\usepackage{amsfonts,amssymb,bm,mathrsfs,bbm,amscd}
\usepackage{lastpage}
\usepackage[pdftex]{graphicx}
\usepackage{multirow}
\DeclareGraphicsExtensions{.pdf,.jpeg,.png}
\usepackage{epstopdf}
\usepackage[english]{babel}
\usepackage[colorlinks=false,pdfpagemode=FullScreen,,setpagesize=off,pdfborder={0 0 0}]{hyperref}
\usepackage{overpic}
\usepackage{lineno}
\usepackage{color}
\usepackage{subfigure}
\usepackage{overpic}
\usepackage{diagbox}
\usepackage{hhline}
\usepackage{multirow}
\usepackage{bigstrut,multirow,rotating}
\usepackage{booktabs}
\usepackage{verbatim} 
\RequirePackage{CJKnumb}

\begin{document}
\begin{CJK*}{UTF8}{gkai}


\title{Prospects of $CP$ violation in $\Lambda$ decay with polarized electron beam at STCF} 

\maketitle

\begin{center}

\author{
	\begin{small}
		\begin{center}
			Sheng Zeng$^{1}$(曾胜),
			Yue Xu$^{2}$(徐月),
			Xiao Rong Zhou$^{3}$$^{,}$\footnotemark[1](周小蓉),
				Jia Jia Qin$^{1}$(秦佳佳),
				Bo Zheng$^{1}$$^{,}$\footnotemark[2](郑波)\\
				\vspace{0.2cm} {\it
					$^{1}$ University of South China, Hengyang 421001, People's Republic of China\\
					$^{2}$ Department of Physics, Liaoning University, Shenyang 110036, People's Republic of China\\
					$^{3}$ School of Physical Sciences, University of Science and Technology of China, Hefei 230026, People's Republic of China\\
				}
			\end{center}
			\vspace{0.4cm}
		\end{small}
	}
	\renewcommand{\thefootnote}{\fnsymbol{footnote}}
	\footnotetext[2]{Corresponding author(1):zhengbo\_usc@163.com}
	\footnotetext[1]{Corresponding author(2):zxrong@ustc.edu.cn}
\end{center}

\begin{abstract}
Based on $1.89 \times 10^8$ $J/\psi \rightarrow \Lambda \bar{\Lambda}$ Monte Carlo (MC) events produced from a longitudinally-polarized electron beam, the sensitivity of $CP$ violation of $\Lambda$ decay is studied with fast simulation software. 
In addition,  the $J/\psi \rightarrow \Lambda \bar{\Lambda}$ decay can also be used as a process to optimize the detector response using the interface provided by the fast simulation software.
In the future, STCF is expected to obtain 3.4 trillion $J/\psi$ events, and the statistical sensitivity of $CP$ violation of $\Lambda$ decay via $J/\psi \rightarrow \Lambda \bar{\Lambda}$ process is expected to reach $\mathcal O$~$(10^{-5})$ when the electron beam polarization is 80\%.
\end{abstract}


\begin{multicols}{2}

\section{Introduction}
The Standard Model (SM) describes elementary particles and their interactions, and is highly consistent with most current experimental results. 
However, it falls short in providing a complete explanation for the observed preponderance of matter over antimatter in the universe.
The study of $CP$ violation provides a crucial perspective on understanding the origin and nature of this fundamental asymmetry, as well as exploring physics beyond the SM~\cite{Sakharov:1967dj, Bernreuther:2002uj,Bediaga:2020qxg}. Currently, $CP$ violation has been discovered in the decay of mesons such as $K$~\cite{Christenson:1964fg}, $B$~\cite{BaBar:2001pki,Belle:2001zzw}, and $D$~\cite{LHCb:2019hro}, but not yet in any baryon decay.
\par In 1956, Lee and Yang firstly proposed the violation of parity ($P$) in weak decays of baryons, and later pointed out that violation of parity implies a violation of charge~($C$) conjugation symmetry as well~\cite{Lee}.
When studying the decays of a spin-1/2 hyperon $\Lambda$ to a final state comprising a spin-1/2 baryon $p$ and a $\pi$, it is observed that the parity-even amplitude results in a $p$-wave state, whereas the parity-odd amplitude leads to an $s$-wave state.~The amplitudes $P_{1}$ and $S_{1}$ are used to denote the parity-even and parity-odd amplitudes, respectively.~The decay parameters of baryons can be represented by the formula:  $\alpha=2\frac{Re(S_{1}*P_{1})}{|S_{1}|^{2}+|P_{1}|^{2}}$. Theoretical physicists Donoghue and Pakvasa proposed that the observable quantity of $CP$ violation could be constructed using asymmetric parameters in the decay of hyperons. They predicted that the $CP$ violation of $\Lambda \rightarrow p \pi^{-}$ in the SM is $\mathcal O$~$(10^{-5})$~\cite{J. F. Donoghue}. As a result, the decay of hyperons is sensitive to sources of $CP$ asymmetry from physics beyond the SM~\cite{Ireland:2019uja}.
In the $CP$-conserving limit, the amplitudes $\bar{S_{1}}$ and $\bar{P_{1}}$ for the charge-conjugated (c.c.)~decay mode of the antibaryon $\bar{\Lambda}\rightarrow \bar{p} \pi^{-}$ are
$\bar{S_{1}}=-S_{1}$ and $\bar{P_{1}}=P_{1}$. Therefore, the decay parameters have the opposite values:~$\alpha_{1}=-\alpha_{2}$. The $CP$ asymmetry can be described as $A_{CP}=\frac{\alpha_{1}+\alpha_{2}}{\alpha_{1}-\alpha_{2}}$. 
 If $CP$ is conserved, then $A_{CP}=0$~\cite{Okubo:1958zza, Pais:1959zza}. 

Experiments specifically aimed at observing hyperon $CP$ violation were conducted by Fermilab E756~\cite{Ho:1991rz} and HyperCP~\cite{HyperCP}. In these experiments, the sum of the observables $A_{CP}^{\Xi^{-}}+A_{CP}^{\Lambda_{p}}$ for $\Xi^{-} \rightarrow \Lambda \pi^{-}$($\Xi^{-}$) and $\Lambda \rightarrow p \pi^{-}$($\Lambda_{p}$) is $0(7)\times10^{-4}$~\cite{HyperCP:2004zvh}. The SM predicted that $A_{CP}^{\Xi^{-}}+A_{CP}^{\Lambda_{p}}$ is $-0.5\times10^{-4} \leq A_{CP}^{\Xi^{-}}+A_{CP}^{\Lambda_{p}} \leq -0.5\times10^{-4}$~\cite{Tandean:2002vy}. In 2019, BESIII adopted an innovative method to study $CP$ violation using hyperon-antihyperon pairs based on $1.3\times 10 ^{9}~J/\psi$ and obtained the most accurate measurement results by then $A_{CP}=-0.006\pm0.012\pm0.010$~\cite{BESIII:2018cnd}. Afterwards, BESIII updated the result using 10 billion $J/\psi$ events,    $A_{CP} = -0.0025 \pm 0.0046 \pm 0.0012$~\cite{BESIII:2022cnd}. Nevertheless, the sensitivity of experiments to $CP$ violation still does not meet theoretical predictions, and currently it is mainly limited by the statistics uncertainty~\cite{Pakvasa:1998nw, Donoghue:1986hh}.
\par The Super Tau-Charm Facility (STCF) is one of the important options of the accelerator-based particle physics large-scale scientific device after the BEPCII collider~\cite{Levichev:2018cvd}. 
The STCF was designed to collect over 1~ab$^{-1}$ of data and has great potential in improving luminosity and realizing beam polarization. It is expected to provide $3.4 \times 10^{12}$ $J/\psi$
events. It is proposed that an electron beam
polarization of 80-90$\%$ at $J/\psi$ energy can be achieved with the same beam current~\cite{I. Koop}. In this paper, we focus on the study of $J/\psi \rightarrow \Lambda\bar{\Lambda}$ and show how the precision of the $CP$ violation of $\Lambda$ decay  in $J/\psi \rightarrow \Lambda\bar{\Lambda} \rightarrow p \bar{p}\pi^{+}\pi^{-}$  is related with the polarization of electron beam.

\section{The STCF detector and Monte Carlo simulation}
Currently, the STCF is in the research and design stage. The center-of-mass energy $(\sqrt{s})$ designed for the STCF ranges from 2 to 7~GeV, with a peak luminosity of at least $0.5\times10^{35}$ cm $^{-2}$s$^{-1}$ or high at $\sqrt{s}=4.0$~GeV. Moreover, luminosity upgrade space will be left and the beam polarization operation of the second phase will be achieved. The STCF will serve as a crucial experiment to test the SM and study potential new physics.
\par In order to manifest the expected high-precision with the high-luminosity samples, the detector design on STCF needs to meet the following requirements: the large coverage angle, high detection efficiency and good resolution of particles result in rapid triggering, and high radiation resistance.
The preliminary design of the STCF detector mainly consists of a tracking system composed of inner and outer trackers, a particle identification (PID) system with $\pi/K$ and $K/\pi$ misidentification less than 2\% with the PID efficiency of $K/\pi$ over 97\%, an electromagnetic calorimeter (EMC) with an excellent energy resolution and a
good position resolution, a super conducting solenoid and
a muon detector (MUD) that provides good $\pi/\mu$ separation. The specific detailed requirements for each subdetector design can be found in the Refs.~\cite{Progress on Preliminary Conceptual study of HIEPA,High Intensity Electron Positron Accelerator}.
\par A fast simulation software, specifically designed for STCF detectors, has been developed to investigate their physical potential and further optimize detector design~\cite{Shi:2020nrf}. Instead of simulating the objects in each subdetector using {\sc Geant4}, the fast simulation models their response, including efficiency, resolution, and other factors used in data analysis, randomly sampling based on the size and shape of their performance. The scaling factor can be adjusted according to the performance limitations of the STCF detector, and these configurations can be easily interfaced. For this analysis, five groups of signal MC samples $J/\psi \rightarrow \Lambda\bar{\Lambda} \rightarrow p \bar{p}\pi^{+}\pi^{-}$ were generated according to the amplitude described in Sec. $\ref{sec:third}$, and were analysed using the fast simulation package to investigate $CP$ violation. Each group consisted of $1.89 \times 10^8$ $J/\psi \rightarrow \Lambda \bar{\Lambda}$ events with a polarization rate of the electron beam ranging from 0 to 1, with a step size of 0.2. 
The obtained signal events is calculated using the following formula: 
\begin{equation}
N_{sig}=N_{J/\psi}\times \mathcal B_{J/\psi \rightarrow \Lambda\bar{\Lambda}}\times \mathcal B_{\Lambda\rightarrow p \pi^{-}} \times \mathcal B_{\bar{\Lambda}\rightarrow \bar{p} \pi^{+}},
	\label{eq0}
\end{equation}
where $N_{sig}$ represents the number of signal events, while $N_{J/\psi}$ represents the total number of $J/\psi$ events as 0.1 trillion. Additionally, $\mathcal B_{J/\psi \rightarrow \Lambda\bar{\Lambda}}$, $\mathcal B_{\Lambda\rightarrow p \pi^{-}}$, and $\mathcal B_{\bar{\Lambda}\rightarrow \bar{p} \pi^{+}}$ denote the branching ratios of $J/\psi \rightarrow \Lambda\bar{\Lambda}$, $\Lambda\rightarrow p \pi^{-}$, and $\bar{\Lambda}\rightarrow \bar{p} \pi^{+}$, to be $1.89 \times 10^{-3}$, 63.9\%, and 63.9\% \cite{Workman:2022ynf}, respectively.

\section{Formalism}\label{sec:third}
In electron-positron collision experiments, the polarization of the beam is reflected in the produced baryon-antibaryon pairs. The helicity frame is defined in Fig.~\ref{figure1}. For the decay $J/\psi \rightarrow \Lambda\bar{\Lambda}$, the $\hat{z}$ axis follows the direction of the positron momentum. The $\hat{z}_{1}$ axis is defined along the momentum vector of the $\Lambda$ baryon, denoted by $\textbf{p}_\Lambda$~=~$-\textbf{p}_{\bar{\Lambda}}$~=~ $\textbf{p}$ in the center-of-mass system of the $e^{+}e^{-}$ collision. The $\hat{y}_{1}$ axis is perpendicular to the production plane and oriented along the vector $\textbf{k}~\times~\textbf{p}$， where $\textbf{k}_{e^{-}}$~=~$-\textbf{k}_{e^{+}}$~=~$\textbf{k}$ is the momentum of electron beam. The scattering angle of the $\Lambda$ is given by $\cos\theta_{\Lambda}$~=~ $\hat{\textbf{p}}$~$\cdot$~$\hat{\textbf{k}}$, where $\hat{\textbf{p}}$ and $\hat{\textbf{k}}$ are unit vectors along the $\textbf{p}$ and $\textbf{k}$ directions.
\par The general expression for the joint density matrix of the $\Lambda\bar{\Lambda}$ pair is~\cite{Perotti:2018wxm}:
\begin{equation}
\rho_{\Lambda\bar{\Lambda}}=\sum\limits_{{\mu\upsilon}=0}^{3}{C}_{\mu\upsilon}\sigma_{\mu}^{\Lambda}\otimes\sigma_{\upsilon}^{\bar{\Lambda}},
\label{eq2}
\end{equation}
where a set of four Pauli matrices $\sigma_{\mu}^{\Lambda}$($\sigma_{\upsilon}^{\bar{\Lambda}}$) in the $\Lambda$($\bar{\Lambda}$) rest frame is used and ${C}_{\mu\upsilon}$ is a $4 \times 4$ real matrix representing polarization and spin correlations of the baryons. The elements of the ${C}_{\mu\upsilon}$ matrix are functions of the production angle $\theta$ of the $\Lambda$ baryon:
\begin{footnotesize}
\begin{equation}
\setlength{\arraycolsep}{1.2pt}
\setlength{\arraycolsep}{1.2pt}
\begin{bmatrix}
\setlength{\arraycolsep}{1.0pt}
1+\alpha_{\psi}\cos^{2}\theta &\gamma_{\psi}P_{e}\sin\theta& \beta_{\psi} \sin \theta \cos \theta &(1+\alpha_{\psi})P_{e}\cos \theta\\\\
\gamma_{\psi}P_{e}\sin\theta & \sin^{2} \theta & 0 &\gamma_{\psi}\sin \theta \cos \theta\\\\
 - \beta_{\psi} \sin\theta \cos\theta& 0 &\alpha_{\psi}\sin^{2}\theta &-\beta_{\psi}P_{e}\sin\theta\\\\
 -(1+\alpha_{\psi})P_{e}\cos\theta&-\gamma_{\psi}\sin \theta\cos\theta&-\beta_{\psi}P_{e}\sin\theta&-\alpha_{\psi}-\cos^{2}\theta
\label{eq3}
\end{bmatrix},
\end{equation}
\end{footnotesize}\\
where $\beta_{\psi}=\sqrt{1-\alpha_{\psi}^{2}} \sin \Delta\Phi$ , $\gamma_{\psi}=\sqrt{1-\alpha_{\psi}^{2}} \cos \Delta \Phi$, $\alpha_{\psi}^{2}+\beta_{\psi}^{2}+\gamma_{\psi}^{2}=1$ and $P_{e}$ is the polarization of the electron beam.
Two factors are naturally connected to the process of the ratio of two helicity amplitudes $\alpha_{\psi}$ and the relative phase of the two helicity amplitudes $\Delta\Phi$ in the real coefficients ${C}_{\mu\upsilon}$ of Eq.~(\ref{eq3}).
The joint angular distribution of the $p/\bar{p}$ pair within the current formalism is described as follows ~\cite{Perotti:2018wxm}:

\begin{align}
 {\rm T_{r}}\rho_{p\bar{p}}\propto\sum\limits_{{\mu\upsilon}=0}^{3}\alpha_{\mu}^{\Lambda}\alpha_{\upsilon}^{\bar{\Lambda}},
 \label{eq5}
\end{align}
where the $\alpha_{\mu}^{\Lambda}(\theta_{1},\varphi_{1},\alpha_{1})$ and $\alpha_{\upsilon}^{\bar{\Lambda}}(\theta_{2},\varphi_{2},\alpha_{2})$ represent the correlation of the spin density matrices in the sequential
decays and the full expressions can be find in~\cite{Perotti:2018wxm} and  $\alpha_{1}(\alpha_{2})$ are the decay asymmetries for $\Lambda \rightarrow p \pi^{-}$($\bar{\Lambda} \rightarrow  \bar{p}\pi^{+}$). 
In the helicity frame of $\Lambda$, $\theta_{1}$ and $\varphi_{1}$ are the spherical coordinates of the $p$ relative to $\Lambda$.
An event of the reaction  $J/\psi \rightarrow \Lambda\bar{\Lambda} \rightarrow p \bar{p}\pi^{+}\pi^{-}$ is specified by the five dimensional vector $\xi=(\theta,\Omega_{1}(\theta_{1}, \varphi_{1}), \Omega_{2}(\theta_{2}, \varphi_{2}))$, and the joint angular distribution $\mathcal{W}(\xi)$ can be expressed as:
\begin{align}
\begin{gathered}
 \mathcal{W}(\xi)= \mathcal{F}_{0}(\xi)+\sqrt{1-\alpha_{\psi}^{2}}\sin(\Delta\Phi)(\alpha_{2}\cdot\mathcal{F}_{3}-\alpha_{1}\cdot\mathcal{F}_{4})\\\hfill
 +\alpha_{1}\alpha_{2}(\mathcal{F}_{1}+\sqrt{1-\alpha_{\psi}^{2}}\cos(\Delta\Phi)\cdot\mathcal{F}_{2}+\alpha_{\psi}\cdot\mathcal{F}_{5})\\\hfill
 +\alpha_{1}\cdot\mathcal{F}_{6}+\alpha_{2}\cdot\mathcal{F}_{7}-\alpha_{1}\alpha_{2}\cdot\mathcal{F}_{8},\hfill
 \label{eq6}
\end{gathered}
\end{align}
where the angular function $\mathcal{F}_{i}(\xi)$~($i$ = 0, 1, ..., 8) are defined as:
\begin{equation}
\begin{aligned}
&\mathcal{F}_{0}(\xi)=1+\alpha_{\psi}\cos^{2}\theta,\\
 &\mathcal{F}_{1}(\xi)=\sin^{2}\theta \sin\theta_{1}\cos\varphi_{1}\sin\theta_{2}\cos\varphi_{2}-\cos\theta^{2}\cos\theta_{1}\cos\theta_{2},\\
 &\mathcal{F}_{2}(\xi)=\sin\theta \cos\theta(\sin\theta_{1}\cos\theta_{2}\cos\varphi_{1}-\cos\theta_{1}\sin\theta_{2}\cos\varphi_{2}),\\
 &\mathcal{F}_{3}(\xi)=\sin\theta \cos\theta \sin\theta_{2}\sin\varphi_{2},\\
 &\mathcal{F}_{4}(\xi)=\sin\theta \cos\theta \sin\theta_{1}\sin\varphi_{1},\\
 &\mathcal{F}_{5}(\xi)=\sin^{2}\theta \sin\theta_{1}\sin\varphi_{1}\sin\theta_{2}\sin\varphi_{2}-\cos\theta_{1}\cos\theta_{2},\\
 &\mathcal{F}_{6}(\xi)=P_{e}(\gamma_{\varphi}\sin\theta \sin\theta_{1}\cos\varphi_{1}-(1+\alpha_{\psi})\cos\theta \cos\theta_{1}),\\
 &\mathcal{F}_{7}(\xi)=P_{e}(\gamma_{\varphi}\sin\theta \sin\theta_{2}\cos\varphi_{2}+(1+\alpha_{\psi})\cos\theta \cos\theta_{2}),\\
 &\mathcal{F}_{8}(\xi)=P_{e}\beta_{\varphi}\sin\theta(\cos\theta_{1}\sin\theta_{2}\sin\varphi_{2}+\sin\theta_{1}\sin\varphi_{1}\cos\theta_{2}).\\
 \label{eq7}
\end{aligned}
\end{equation}
\begin{center}
	\centering
	\includegraphics[width=0.45\textwidth]{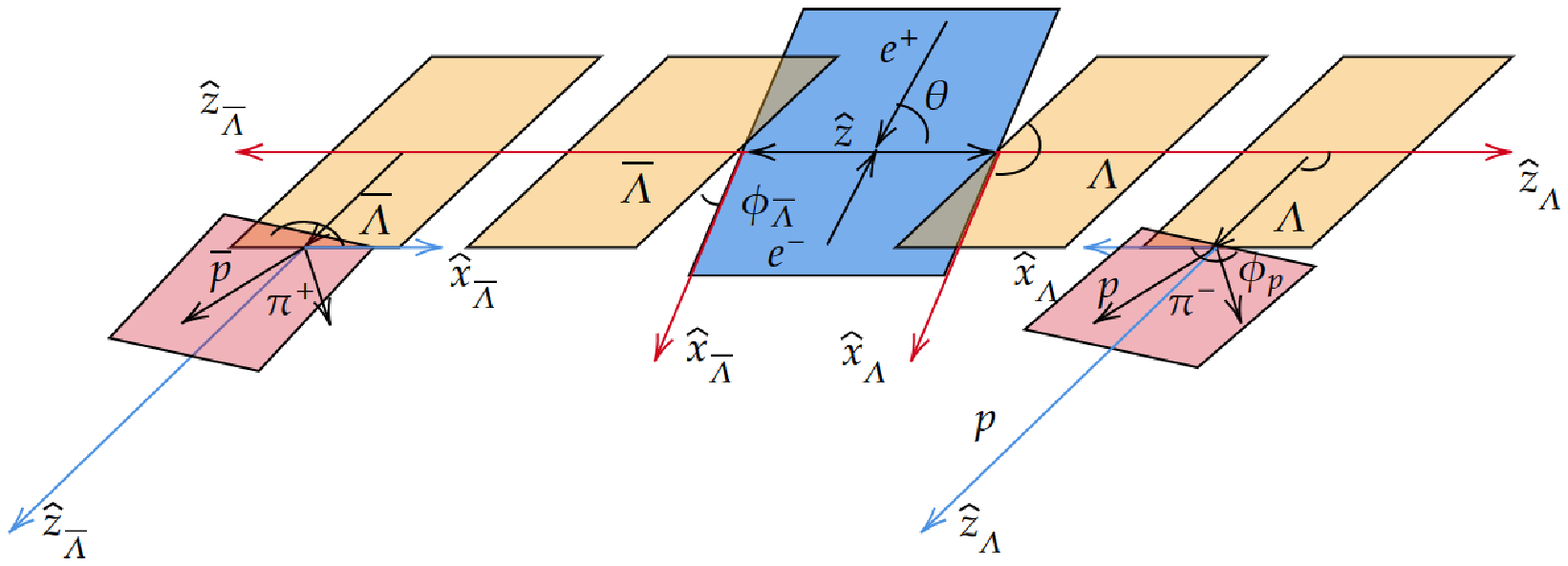}
\figcaption{\label{FIG}Definition of coordinate system used to describe the $e^{+} e^{-} \rightarrow \Lambda\bar{\Lambda} \to p \bar{p}\pi^{+}\pi^{-}$.}
	\label{figure1}
\end{center}  
The Eq.~(\ref{eq7}) contains four terms: $\mathcal{F}_{0}$ describes the angular distribution of $\Lambda$, while $\mathcal{F}_{3}$ and $\mathcal{F}_{4}$ account for the transverse polarization of $\Lambda$ and $\bar{\Lambda}$, respectively. The spin correlations between the two hyperons are described by $\mathcal{F}_{1}$, $\mathcal{F}_{2}$ and $\mathcal{F}_{5}$. 
The terms $\mathcal{F}_{6}$, $\mathcal{F}_{7}$, and $\mathcal{F}_{8}$ describe the beam polarization. In Eq.~(\ref{eq6}), the values of $\alpha_{1}$, $\alpha_{2}$, $\alpha_{\psi}$, and $\Delta\Phi$ are all referenced from Ref.~\cite{BESIII:2022cnd}. The $\Lambda$  polarization
vector $\mathbf{P_{\Lambda}}$ is defined in the rest frame of $\Lambda$ as shown in Ref.~\cite{Salone:2022lpt}:
\begin{align}
\mathbf{P_{\Lambda}}=\frac{\gamma_{\psi} P_{e}\sin\theta\hat{x}_{1}-\beta_{\psi} \sin\theta \cos\theta\hat{y}_{1}-(1+\alpha_{\psi})P_{e}\cos\theta\hat{z}_{1}}{1+\alpha_{\psi}\cos^{2}\theta}.
\label{eq8}
\end{align}
The distribution of the module of $\left|\mathbf{P_{\Lambda}}\right|$ versus the polar angle of $\Lambda$ is determined as shown in Fig.~\ref{figure2}.

\begin{center} 
    \includegraphics[width=0.45\textwidth]{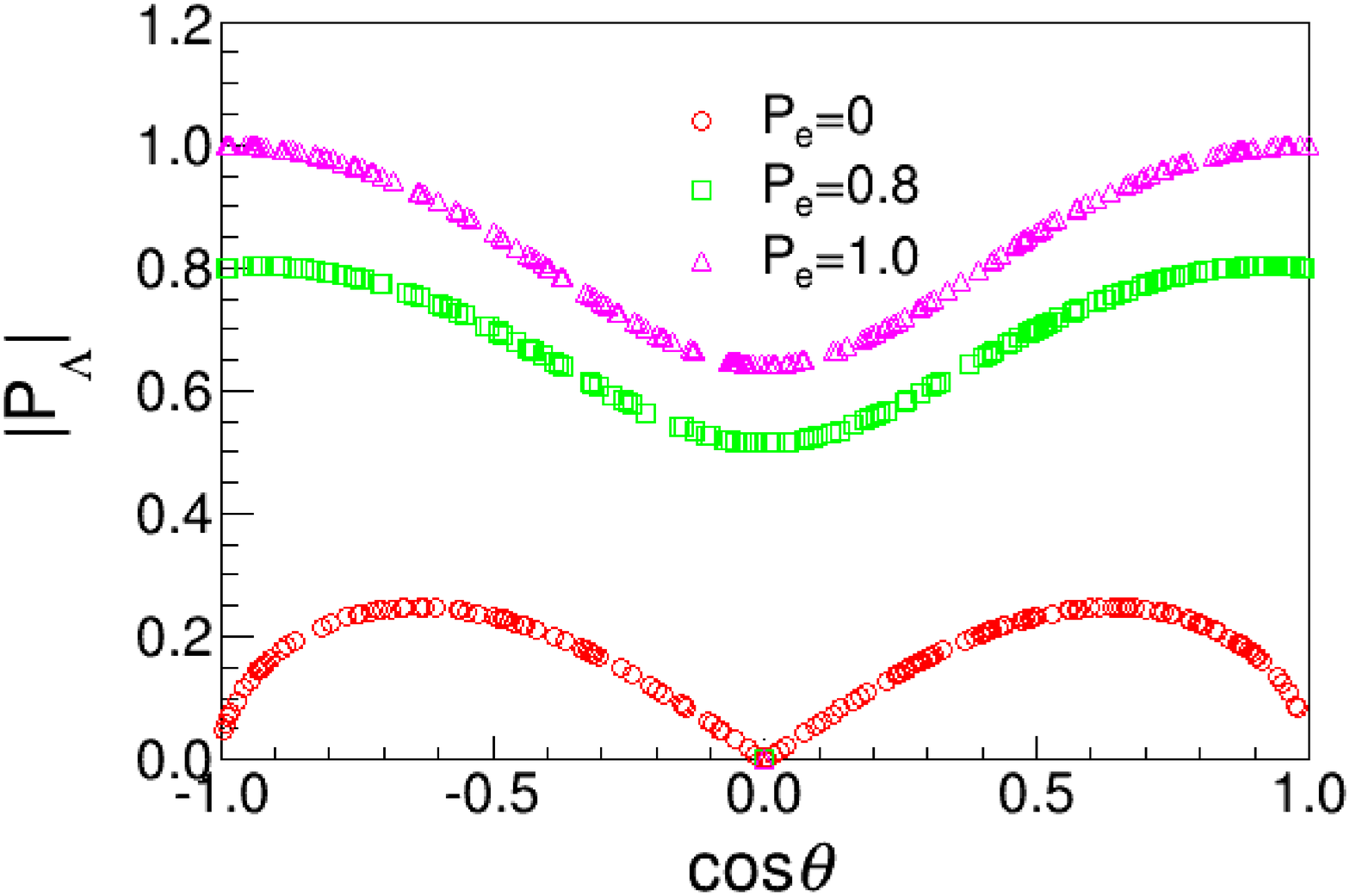}
	\figcaption{ \label{figure2} Distribution of $\mathbf{P_{\Lambda}}$ versus $\cos\theta$. The red circle, green square, pink triangle represent the three cases of electron beam polarization 0, 0.8 and 1.0 respectively.}
    \label{figure2}
\end{center}
The relation between the electron beam polarization and the polarization of $\Lambda$ can be described by the following formula \cite{Salone:2022lpt}:
\begin{equation}
	\begin{aligned}
		\left \langle \mathbf{P_{\Lambda}} \right \rangle=\frac{(1-\alpha_{\psi}^{2})\sin^{2}(\Delta\Phi)}{\alpha_{\psi}^{2}(3+\alpha_{\psi})}(3+2\alpha_{\psi}-3(1+\alpha_{\psi})\frac{\arctan\sqrt{\alpha_{\psi}}}{\sqrt{\alpha_{\psi}}})+\\
    \frac{3(1+\alpha_{\psi})^{2}}{\alpha_{\psi}(3+\alpha_{\psi})}(1-\frac{1-\alpha_{\psi}}{1+\alpha_{\psi}}\cos^{2}(\Delta\Phi)\frac{\arctan\sqrt{\alpha_{\psi}}}{\sqrt{\alpha_{\psi}}})P^{2}_{e}
	\label{eq_PB}.
\end{aligned}
\end{equation}

Figure~\ref{PB} provides an intuitive representation of the relation between the polarization of $\Lambda$ and the polarization of electron beam.
\begin{center} 
	\includegraphics[width=0.45\textwidth]{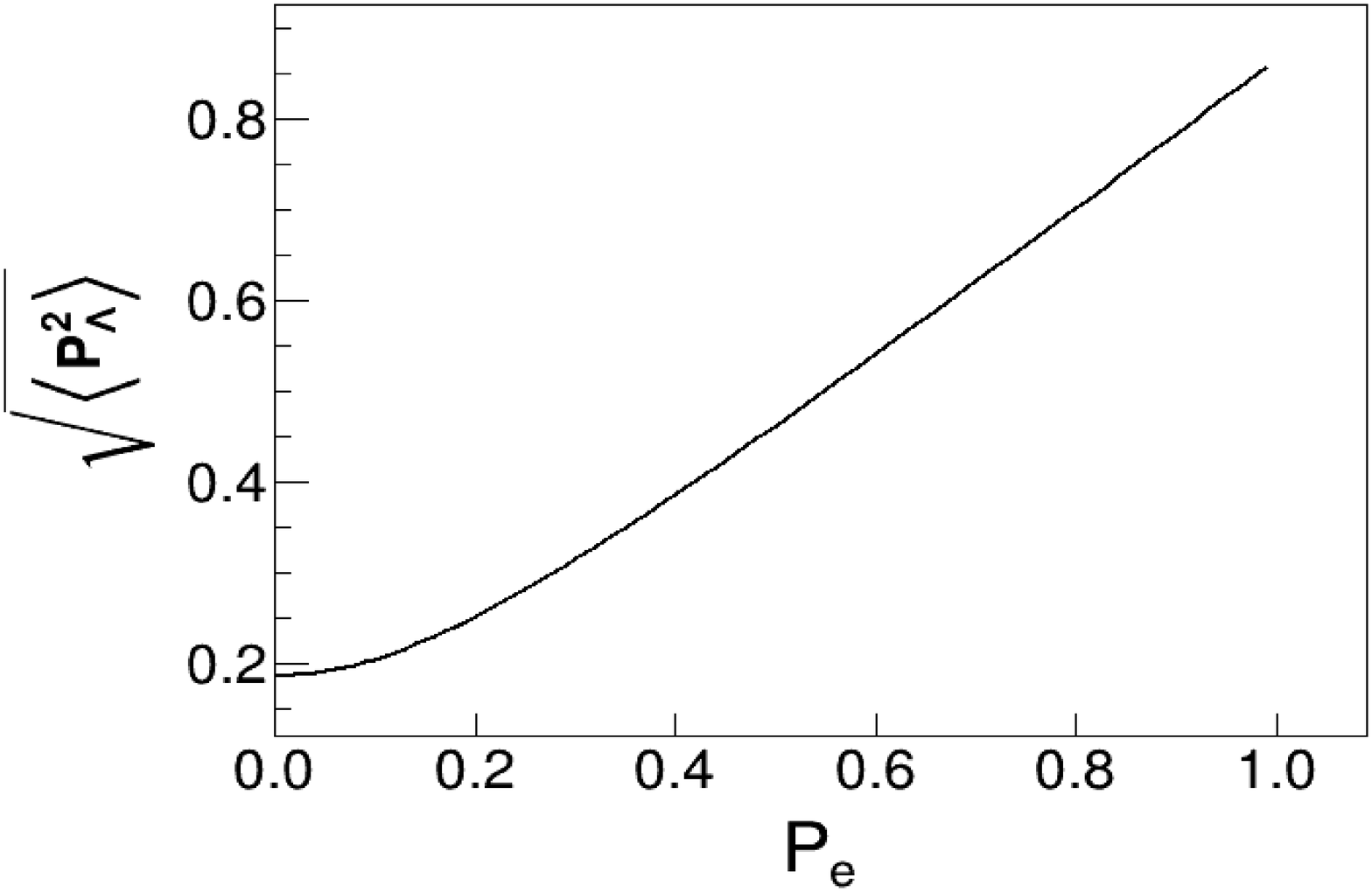}
	\figcaption{The relation between the polarization of $\Lambda$ and the polarization of electron beam.}
	\label{PB}
\end{center}
\section{Analysis of $J/\psi \rightarrow \Lambda\bar{\Lambda} \rightarrow p \bar{p}\pi^{+}\pi^{-}$ with fast simulation}
Charged tracks are selected based on the criteria in the fast simulation. Efficiency loss occurs due to the acceptance requirement $|\cos\theta|~\textless~0.93$, where $\theta$ is set in reference to the beam direction, and the requirements on the $\Lambda$ mass and decay vertex.
The decay process of $J/\psi \rightarrow \Lambda\bar{\Lambda}$ can be described as follows: the $\Lambda$ decays into a $p$ and $\pi^{-}$, while the $\bar{\Lambda}$ decays into a $\bar{p}$ and $\pi^{+}$. Therefore, the candidate event must have at least four charged tracks.
Charged tracks are divided into two categories, where positively charged tracks are $p$ and $\pi^{+}$, and negatively charged tracks are $\bar{p}$ and $\pi^{-}$.
Based on the momentum of tracks, the momentum of the particle track can subsequently be identified as $p (\bar{p})$ or $\pi^{+} (\pi^{-})$. Specifically, particles with a momentum exceeding 500~MeV/$c$ are identified as $p (\bar{p})$, while those with a momentum less than 500~MeV/$c$ are classified as $\pi^{+} (\pi^{-})$. For selected events, multiple charged tracks must be present for $p$, $\bar{p}$, $\pi^{+}$ and $\pi^{-}$, respectively.
A second vertex fit is performed by looping over all combinations of positive and
negative charged tracks. The $p$($\bar{p}$) $\pi^{-}$($\pi^{+}$) pairs selected must decay from the same vertex. The invariant mass of $\Lambda$($\bar{\Lambda}$) must fall within the range of $1.1107<M_{p \pi^{+}}<1.1207$ GeV/$c^{2}$($1.1107<M_{\bar{p} \pi^{-}}<1.1207$ GeV/$c^{2}$). After applying the aforementioned selection criteria, the spectrum of the energy invariant mass of $p$ and $\pi^{-}$ from
$\Lambda$ in the mass frame is shown in Fig.~\ref{figure3}, and corresponds to about 3.5 times mass resolution of $\Lambda$. The event selection efficiency is finally determined to be 38.2\%.   
The influence of the polarization of electron beam on event selection efficiency has also been studied, and the selection efficiency is found to be unaffected by beam polarization.
\begin{center}
    \includegraphics[width=0.45\textwidth]{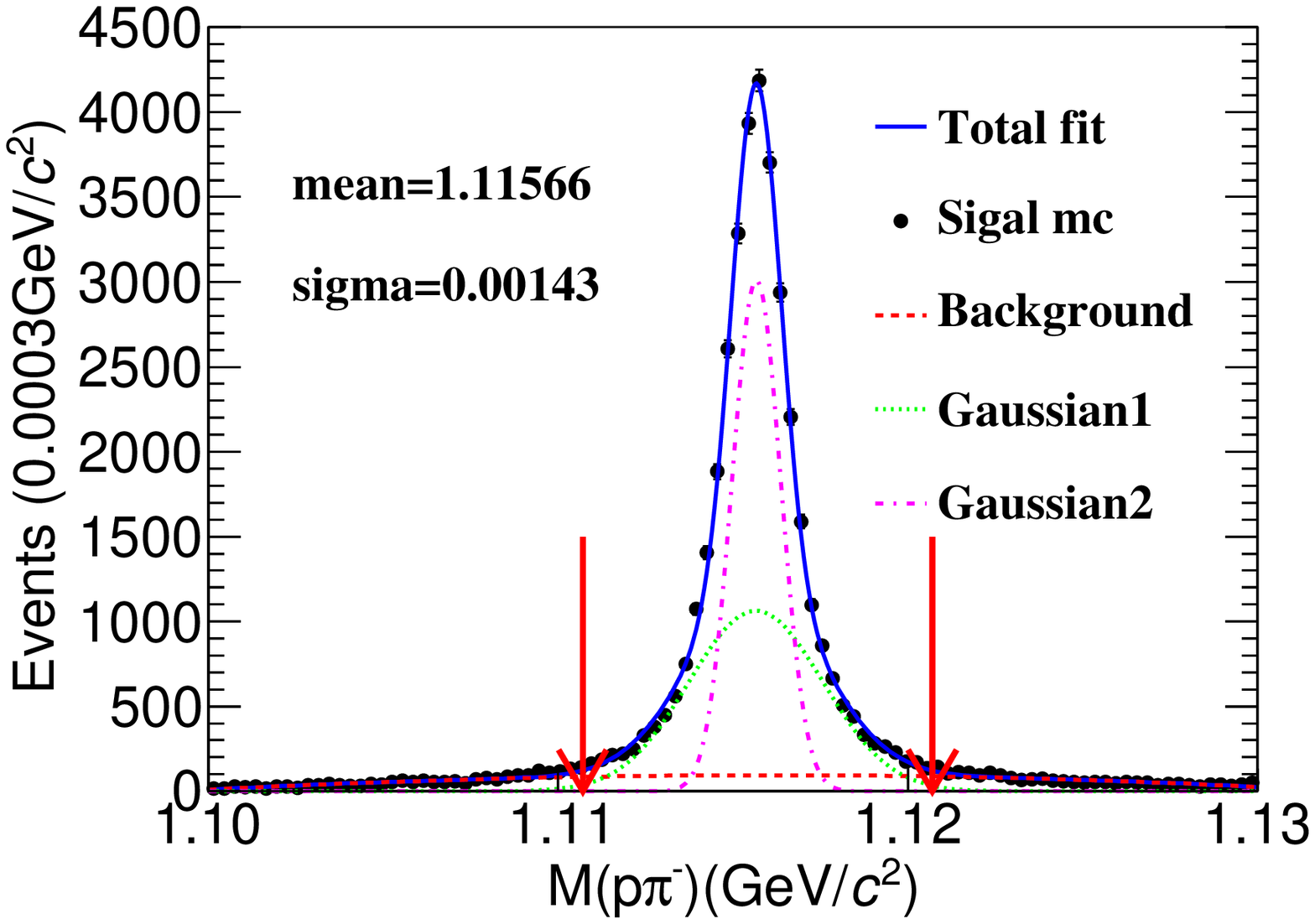}
	\figcaption{ \label{Figure2}The spectrum of the invariant mass of the $p$ and $\pi^{-}$ from in the mass frame of $\Lambda$. The black dots with error bars represent the signal MC sample. The blue solid curve represents the fit results, where the signal function is a double Gaussian consisting of a pink dashed line and a green dashed line, respectively. The background is represented by the solid red line, which is modeled using a second-order polynomial.}
    \label{figure3}
\end{center}
\section{Optimization of detector response}
 To enhance the performance of the detector, the momentum resolution and the selection efficiency of the charged tracks can be further optimized with the help of fast simulation software. The optimization results are presented as follows:
\begin{itemize}
\item Tracking efficiency 
\par The detector is capable of identifying the following charged particles in the final state: electron, muon, pion, kaon and proton. It is required to cover a wide momentum range while maintaining a high reconstruction efficiency throughout this range. To further enhance the capability of reconstructing low-momentum tracks, different materials or sophisticated tracking algorithms can be employed to further enhance the capability of low momentum tracks reconstruction in the design portion of the STCF tracking system. Low momentum final-state particles $\pi$ mesons are produced by the decay of the $J/\psi \rightarrow \Lambda\bar{\Lambda} \rightarrow p \bar{p}\pi^{+}\pi^{-}$, which is a useful option for enhancing the resolution of low momentum particles and optimizing detector response.
In this analysis, the tracking efficiency scaling factor was gradually increased from 1.0 to 2.0. The scale factor represents the ratio of the efficiency. Figure~\ref{figure4} shows that the final selection efficiency greatly improves between 1.0 and 1.1, resulting in an improvement form 38.2\% to 43.7\%.
\begin{center}
	\centering
   \includegraphics[width=0.45\textwidth]{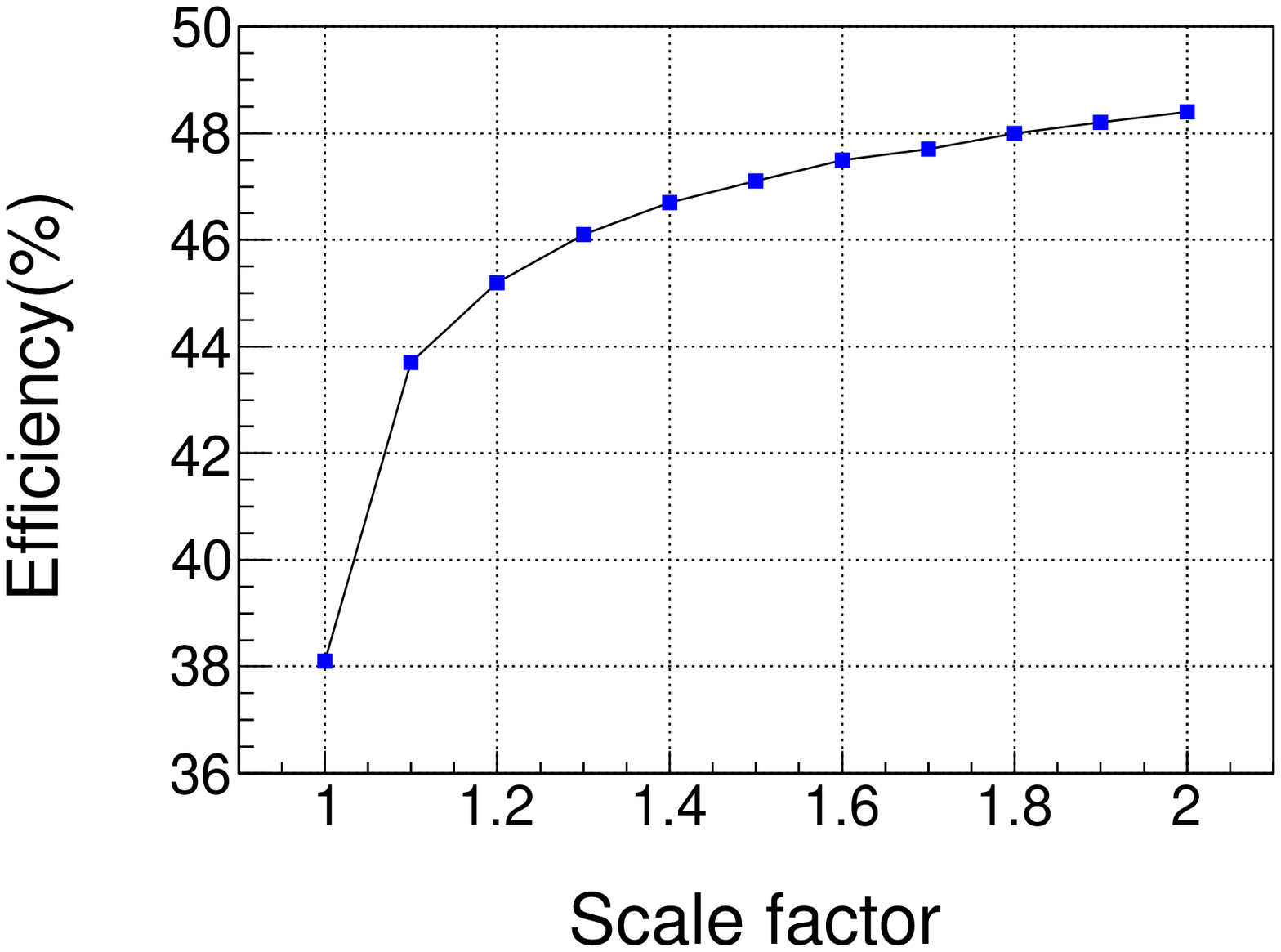}
	\figcaption{\label{figure4}Scales of charged track efficiency versus selection efficiency.}
\end{center}

\item Momentum resolution of tracking 
\par In the fast simulation, the momentum resolution of the charged track can also be adjusted for improvement. The resolution of tracking system in the $xy$ plane and $z$ direction are $\sigma_{xy}$ and $\sigma_{z}$, respectively. The default values of $\sigma_{xy}$ and $\sigma_{z}$ are 130~$\mu$m and 2480~$\mu$m, respectively. Optimization is performed on $\sigma_{xy}$ from 0~$\mu$m to 130~$\mu$m, and on $\sigma_{z}$ from 0~$\mu$m to 2480~$\mu$m,  resulting in no significant change in detection efficiency, as demonstrated in Fig.~\ref{figure5}.
\begin{center}
	\centering
   \includegraphics[width=0.45\textwidth]{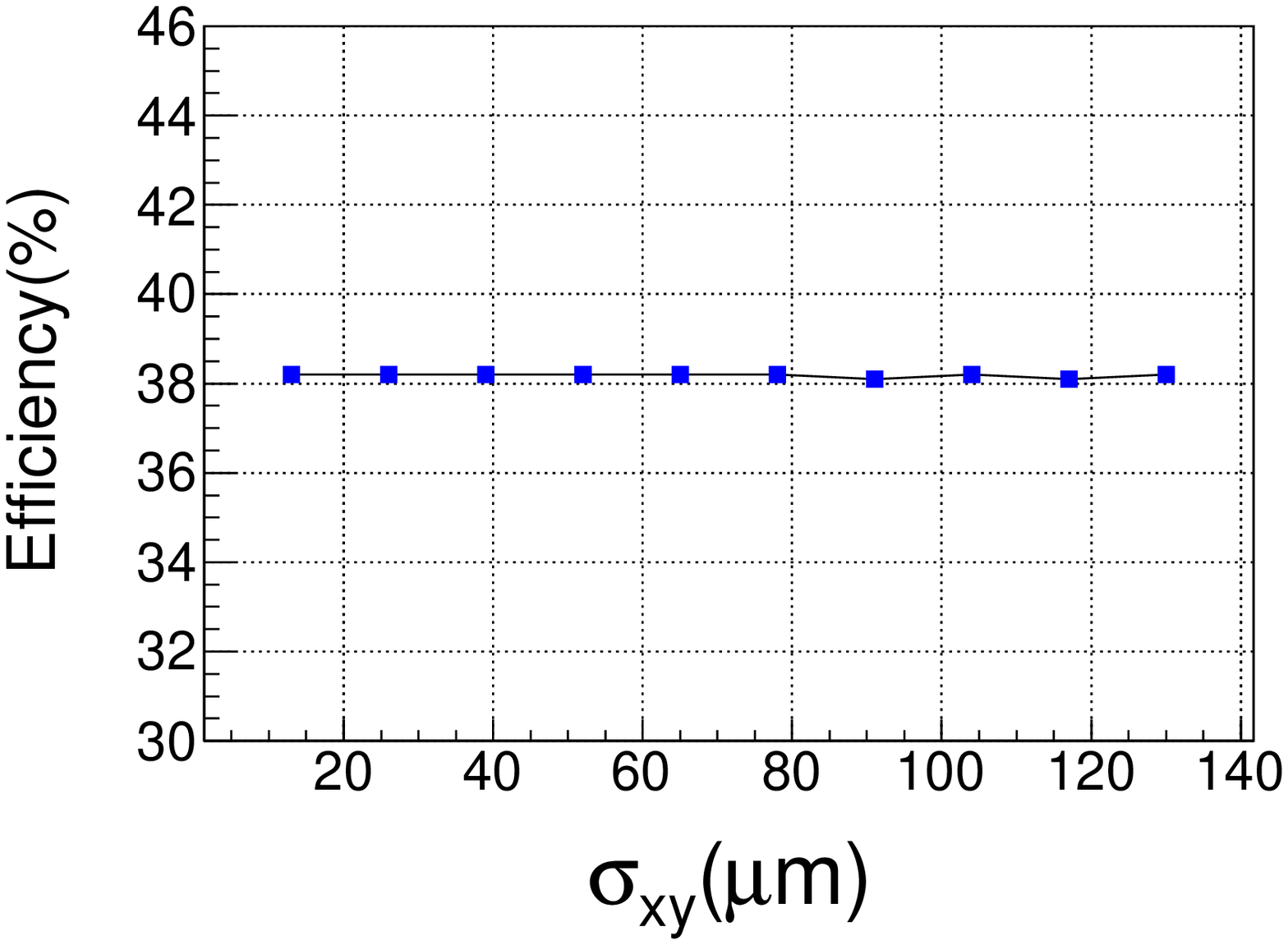}
	\figcaption{\label{figure5}Momentum resolution of charged tracks versus the selection efficiency.}
\end{center}
\end{itemize}

\section{Angular distribution fitting}
Based on the joint angular distribution, a maximum likelihood fit is performed with four free parameters ($\alpha_{\psi}$,$\alpha_{1}$,$\alpha_{2}$,$\Delta\Phi$). The joint likelihood function, as defined in reference~\cite{MLL} and shown in Eq.~(\ref{eq9}) , is used for this purpose.
\begin{footnotesize}
	\begin{align}
		\mathcal{L}=\prod\limits_{i=1}^{N}\mathcal{P}(\xi^{i},\alpha_{\psi},\alpha_{1},\alpha_{2},\Delta\Phi)=\prod\limits_{i=1}^{N}\mathcal{C}\mathcal{W}(\xi^{i},\alpha_{\psi},\alpha_{1},\alpha_{2},\Delta\Phi)\epsilon(\xi^{i}),
		\label{eq9}
	\end{align}
\end{footnotesize}\\
The probability density function of the kinematic variable $\xi^{i}$ for event $i$ denoted as $\mathcal{P}(\xi^{i}, \alpha_{\psi}, \alpha_{1}, \alpha_{2}, \Delta\Phi)$ is used in a maximum likelihood fit. The fit is performed using Eq.~(\ref{eq7}), where $\mathcal{W}(\xi^{i}, \alpha_{\psi}, \alpha_{1}, \alpha_{2}, \Delta\Phi)$
represents the weights assigned to each event. The detection efficiency is represented by $\epsilon(\xi^{i})$ and $N$ denotes the total number of events. The normalization factor, denoted as $\mathcal{C}^{-1} =\frac{1}{N_{MC}}\sum\limits^{N_{MC}}_{j=1}\mathcal{W}(\xi^{j}, \alpha_{\psi}, \alpha_{1}, \alpha_{2}, \Delta\Phi)\epsilon(\xi^{j})$ is estimated the $N_{\rm MC}$ events generated with the phase space model, which is about ten times the size of mDIY MC. Usually, the minimization of $-{\rm ln}\mathcal{L}$ is performed by using MINUIT~\cite{MINUIT}:
\begin{align}
	-{\rm ln}\mathcal{L}=-\sum\limits^{N}_{i=1}{\rm ln}\mathcal{C}\mathcal{W}(\xi^{i},\alpha_{\psi},\alpha_{1},\alpha_{2},\Delta\Phi)\epsilon(\xi^{i})
	\label{eq10}.
\end{align} 
In this analysis, we extrapolate the sensitivity of $CP$ violation for a large number of $J/\psi$ events generated at future STCF, taking into account the effects of excessive disk pressure. This extrapolation is based on the relationship between the sensitivity of 
$CP$ violation and generated 0.1 trillion $J/\psi$ events. We investigate this relation using a sample size ranging from 0.01 to 0.1 trillion 
$J/\psi$, with a step size of 0.01 trillion 
$J/\psi$. The sensitivity analysis is presented in Fig.~\ref{figure6}, where we examine the impact of event statistics on the sensitivity of 
$CP$ violation. The sensitivity can be described using the following formula: 
\begin{equation}
	\sigma_{A_{CP}}\times\sqrt{N_{fin}}=k.
	\label{ex}
\end{equation}
\begin{center}
	\centering
	\includegraphics[width=0.45\textwidth]{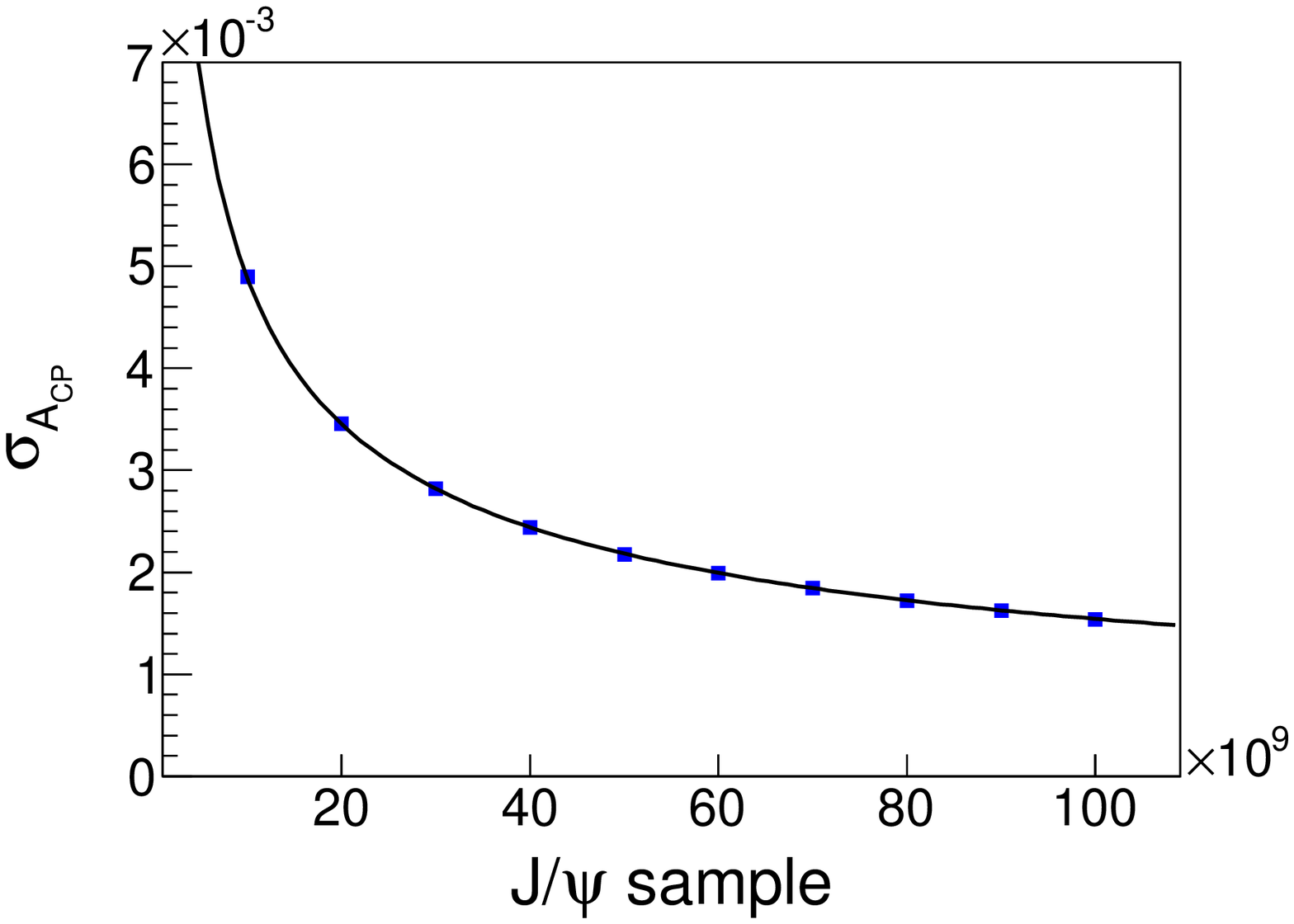}
	\figcaption{\label{figure6}The blue dots represent the statistical errors of $CP$ violation, while the black line is fitted using Eq.~(\ref{ex}).}
\end{center}
The variable $N_{fin}$ represents the number of events that pass the final selection criteria, while $k$ is a constant with a value of 7.82.
\par As shown in Fig.~\ref{figure6}, the sensitivity of statistical errors increases proportionally with the square root of the number of signal events. This observation provides a foundation for extrapolating 
$CP$ violation sensitivity from the size of the data sample.
\par Five different beam polarizations were utilized to generate a sample of 0.1 trillion MC events, with the specific aim of investigating the quantity 
$\sigma_{A_{CP}}$. The resulting five sets of data points were fit using Equation~(\ref{eq11}) \cite{Salone:2022lpt}: 
\begin{equation}
	\sigma_{A_{CP}}\approx \sqrt{\frac{3}{2}} \frac{1}{\alpha_{1}\sqrt{N_{sig}}\sqrt{\left \langle P_{B}^{2} \right \rangle}}.
	\label{eq11}
\end{equation}

\begin{center}
	\centering
	\includegraphics[width=0.45\textwidth]{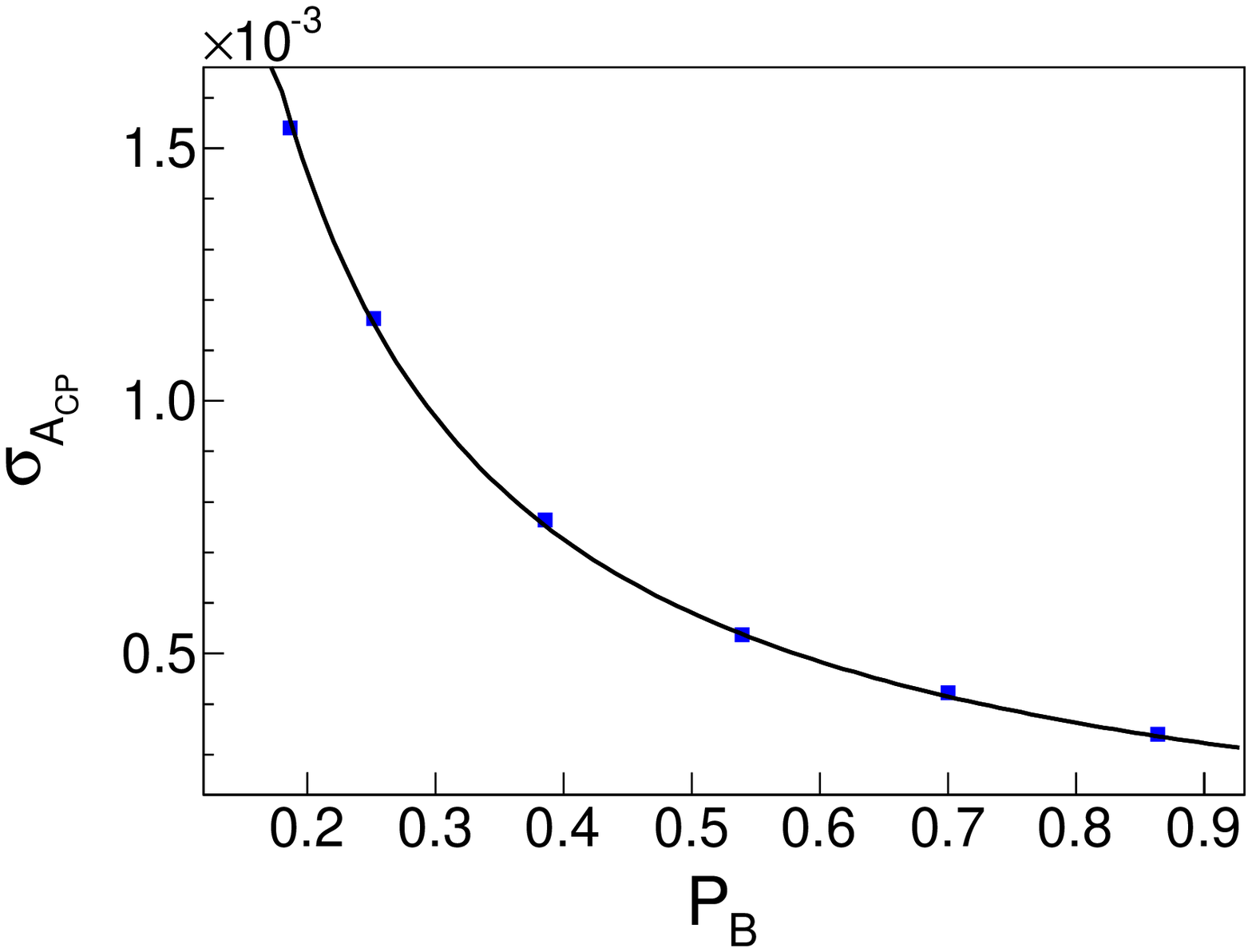}
	\figcaption{\label{figure7}The blue dots represent the values of $\sigma_{A_{CP}}$ under different $\Lambda$ polarization.}
\end{center}

By extrapolating the number of $J/\psi$ events based on the 3.4 trillion events expected to be generated annually by future STCF, as demonstrated in Fig.~\ref{figure6} and Fig.~\ref{figure7}, it is apparent that the statistical sensitivity of 
$CP$ violation will reach the order of $\mathcal O$ $(10^{-5})$ at a beam polarization of 80\%.

\section{Summary and prospect}
In this paper, the detection efficiency and $CP$ violation sensitivity of the $J/\psi \rightarrow \Lambda\bar{\Lambda} \rightarrow p \bar{p}\pi^{+}\pi^{-}$ process under different beam polarization are studied by using the $1.89 \times 10^8$ $J/\psi \rightarrow \Lambda \bar{\Lambda}$ MC samples generated by the fast simulation package developed during the pre-research stage of STCF.
We find that the polarization of the electron beam does not affect the final detection efficiency, but the sensitivity of $CP$ violation increases with increasing electron beam polarization. Moreover, if the tracking efficiency of charged particles with low momentum can be improved by 10\% compared to BESIII, the final detection efficiency will significantly improve by 14.3\%. This places high demands on each sub-detector in the expected STCF design. If the future STCF can achieve significant luminosity improvement and beam polarization, the sensitivity of $CP$ violation will be greatly enhanced, making it an ideal place to test $CP$ violation in the SM.

\section{Acknowledgments}
We thank the Hefei Comprehensive National Science Center for their strong support on the STCF key technology research project. 
This work is supported by the National Key R\&D Program of China under Contract No. 2022YFA1602200, the science and technology innovation Program of Hunan Province, No.  2020RC3054 and the international partnership program of the Chinese Academy of Sciences Grant No. 211134KYSB20200057.
\end{multicols}

\vspace{-1mm}
\centerline{\rule{80mm}{0.1pt}}
\vspace{2mm}

\begin{multicols}{2}

\end{multicols}

\clearpage
\end{CJK*}
\end{document}